\documentclass[10pt,letterpaper]{article}
\usepackage{osajnl2}
\usepackage{cite}

\begin{document}

\title{Normalized ghost imaging}

\author{Baoqing Sun,*$^1$ Stephen S.~Welsh,$^1$ Matthew~P.~Edgar,$^1$ Jeffrey H. Shapiro,$^2$ and Miles~J.~Padgett$^1$}
\address{$^1$School of Physics and Astronomy, SUPA, University of Glasgow, G12 8QQ, UK}
\address{$^2$Research Laboratory of Electronics, Massachusetts Institute of Technology, Cambridge, Massachusetts 02139, USA}

\email{*sunbaoqing727@gmail.com} 

\homepage{www.gla.ac.uk/schools/physics/research/groups/optics/}

\begin{abstract}
We present an experimental comparison between different iterative ghost imaging algorithms. Our experimental setup utilizes a spatial light modulator for generating known random light fields to illuminate a partially-transmissive object. We adapt the weighting factor used in the traditional ghost imaging algorithm to account for changes in the efficiency of the generated light field. We show that our normalized weighting algorithm can match the performance of differential ghost imaging. 
\end{abstract}


%\bibliographystyle{ieeetr}
%\bibliography{ref2}

\addcontentsline{toc}{chapter}{Bibliography}

\section{Introduction}

Classical ghost imaging (GI) \cite{Ryan2002PRL, Gatti2004PRL, Gatti2004PRA, Valencia2005PRL, Ferri2005PRL} uses a series of random light patterns to illuminate an unknown object. For each pattern the reflected or transmitted light is measured using a single element detector. The series of single element measurements, combined with the known light patterns is used to deduce the object. In some systems the random light pattern is produced as a time varying laser speckle, and a beam splitter is used to illuminate both the unknown object and a reference camera, with which the pattern is recorded. Subsequently, the need for the beam splitter and camera has been removed by implementing a spatial light modulator (SLM) to produce a random, but known, pattern thereby reducing the number of components in the system necessary for GI experiments \cite{Shapiro2008PRA,Bromberg2009PRA}. This latter approach is known as computational GI and in terms of the experimental arrangement is closely related to the field of single pixel cameras \cite{Duarte2008}.

In all approaches to GI an algorithm is employed to deduce the object using the series of measurements from the single element detector and either the recorded or computationally predicted random patterns. The algorithms employed fall into two categories, iterative ones that give a refined estimate of the object after every new light pattern and measurement, and inversion ones which infer an object based on the entire series of patterns and measurements.

Iterative algorithms use the measured signal to derive a weighting factor to the corresponding pattern that is then added to the iterative estimate of the object. In this paper we compare a number of these iterative algorithms within the context of computational GI. The algorithms we consider are traditional GI (TGI) and differential GI (DGI) \cite{Ferri2010PRL}. In a computational GI setup, TGI uses a weighting factor equal to the signal from the detector whereas DGI utilizes a weighting factor that depends on fluctuations in the measured signal and uses an additional detector to give a normalization. Beyond these two algorithms we introduce a variant of the TGI algorithm, normalized GI (NGI), which we show can match the performance of DGI. 

Key to all these algorithms is that the changes in the measured signal should arise from the overlap of the known random pattern with the unknown object. Obviously other sources of signal change are possible; including fluctuations arising from changes in the source intensity and changes in the efficiency with which the pattern is imprinted. These later sources of noise scale with the signal level and hence become more significant when the signal is high.

\section{Experimental setup}
\begin{figure}[htbp]
\centering\includegraphics[width=9cm]{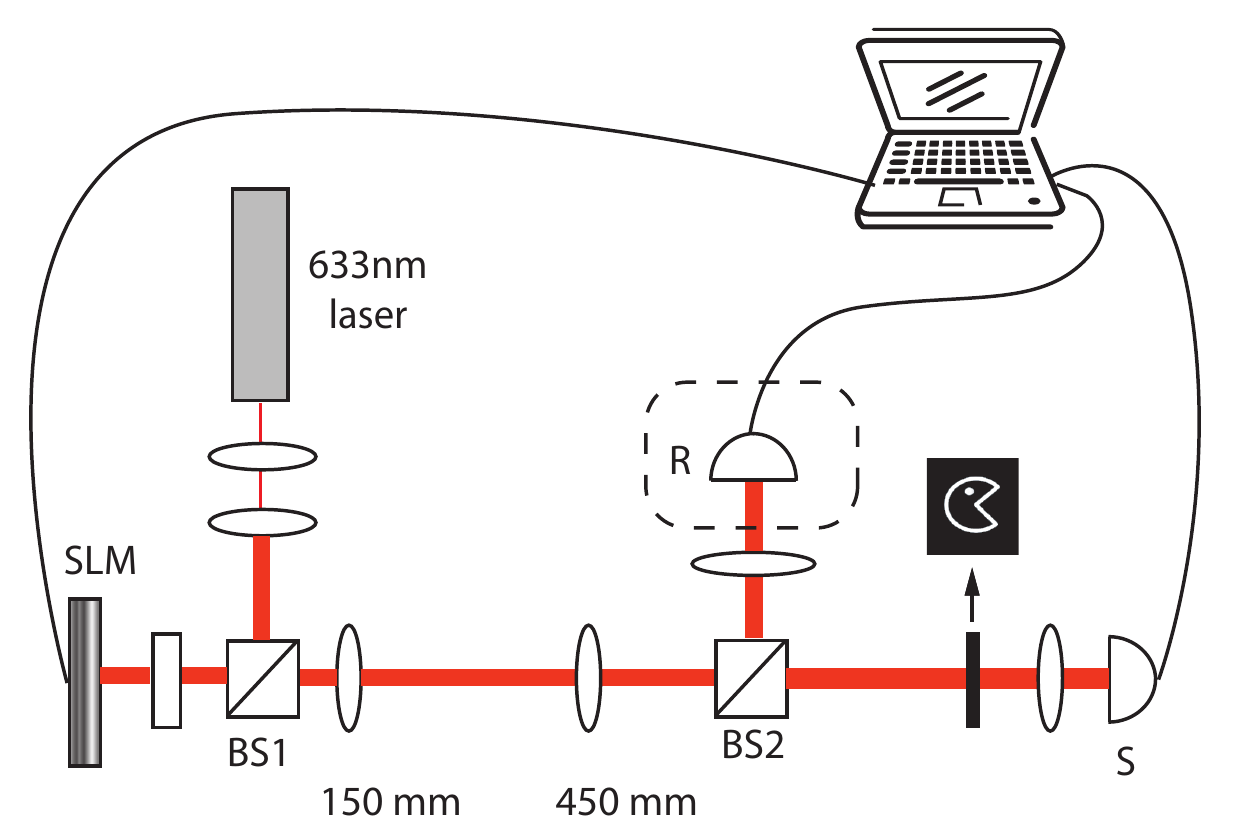}
\caption{Computational ghost imaging setup used in the experiment. A spatial light modulator (SLM) is used to generate a random speckle field, as described in the text and a beam splitter (BS) is used to measure a reference signal R on a bucket detector before the object. The signal, S, is measured on a bucket detector which collects the light transmitted after the object.}
\label{ExperimentalSetup}
\end{figure}

The experimental setup is shown in Figure \ref{ExperimentalSetup}. Here a random light pattern is generated from a simulated superposition of plane waves using random numbers, which is then sent to an SLM to produce a synthesized speckle field. The SLM has $512\times512$ pixels in the window of size $3.584\times3.584\,{\rm mm}$. We pass a collimated laser of wavelength $\lambda= 632.8\,{\rm nm}$ through a polarizing beam splitter and a half-wave plate, before illuminating the SLM window. The speckle field is generated by modulation of the SLM and the returning light field is then magnified by a simple telescope system consisting of $150\,{\rm mm}$ and $450\,{\rm mm}$ biconvex lenses. The object is located at the focus plane of the $450\,{\rm mm}$ lens, which is also the image plane of the SLM window. A $50:50$ beam splitter is placed before the object in order to split the speckle field into two beams; the object beam ($I(x_S)$) and the reference beam ($I(x_R)$). The object beam illuminates the object and is then collected by a bucket detector, thus providing an computational GI setup. The additional reference beam for monitoring the light differentiates our system from previous experimental computational GI configurations. Since we are generating a computer hologram that is then sent to the SLM to create the speckle field, we can therefore predict the light field at the reference arm, negating the demand for a CCD camera, and requiring only a second bucket detector. It should be noted that for TGI based on our computational GI setup, only the object bucket detector is needed. The additional bucket detector in the reference arm is only required for NGI and DGI. Light intensities detected by the object and reference bucket detectors are indicated by $S$ and $R$ respectively, and the speckle field is described by $I(x,y)$. As we use a $50:50$ beam splitter, it is understood that $I(x,y)=2 I(x_S,y_S)= 2I(x_R,y_R)$.

\section{Iterative ghost imaging algorithms}
In all iterative GI techniques, the transmitting object located after the beam splitter, ${\rm BS_2}$, is reconstructed by correlating the speckle field intensity measured at $S$ and $R$, then adding together each successive frame with a suitable weighting factor. The transmitted light power detected after the object can be expressed as
\begin{equation}
\label{S}
S = \int_{A_{l}}I(x_S,y_S)T(x_S,y_S)dx_S dy_S,
\end{equation}
where the laser area is $A_{l}$ and $T(x_S,y_S)$ is the (intensity) object transmission function, while the background reference is expressed as
\begin{equation}
\label{R}
R = \int I(x_R,y_R)dx_R dy_R.
\end{equation}
\subsection{Traditional Ghost Imaging}
In TGI, the reconstruction result of the object, $O(x,y)$ is retrieved from the correlation between $S$ and $I(x,y)$. We define for each iteration, $i$, the contribution to the reconstruction to be \cite{Bromberg2009PRA}
\begin{equation}
\label{TGIi}
O_i(x,y) = \left(S-\left<S\right>\right)\left(I(x,y)-\left<I(x,y)\right>\right),
\end{equation}
where $<.> \equiv \frac{1}{M}\Sigma_r$ denotes an ensemble average for $M$ iterations. We obtain the final reconstruction by averaging over all iterations such that $O(x,y) = \left<O_i(x,y)\right>$. It is easy to understand the reconstruction as being derived from the weighted sum of the speckle field for each measurement. Therefore $S$ is the weight for the speckle field for each measurement. One drawback of using this algorithm is that the reconstruction is heavily weighted to the size of the signal $S$ and is thus susceptible to fluctuations in the generated light field. These fluctuations can arise from either changes to the laser power or the efficiency of the SLM in computational GI.

\subsection{Differential ghost imaging}
Differential GI \cite{Ferri2010PRL}, first performed by Ferri {\it et al}, utilizes a second bucket detector to extract a reference signal which is used in the reconstruction to weight the speckle field based on the average transmission signal relative to the average reference signal. Similarly, each contribution to the reconstruction can be expressed as
\begin{equation}
O_i(x,y) = \left(S - \frac{\left<S\right>}{\left<R\right>}R\right)\left(I(x,y)-\left<I(x,y)\right>\right).
\label{DGIi}
\end{equation}
Thus we obtain the final result by summing for all iterations. We observe the second term in brackets on the right hand side of Eq.\,\ref{TGIi} and Eq.\,\ref{DGIi} are both identical however the first term in brackets of Eq.\,\ref{DGIi} is now weighted according to the average value of $S$, which is normalized to the average value of $R$. As demonstrated in \cite{Ferri2010PRL} the DGI algorithm improves by order of magnitude the SNR of the measurement with respect to TGI. Moreover, a key difference from TGI, it is no longer sensitive to other sources of noise. For example, fluctuations in the laser power or changes to the SLM efficiency will affect both the reference signal and the transmitted signal, and thus the contribution to the reconstruction will be weighted more appropriately. 

\subsection{Normalized ghost imaging}
\subsubsection{Normalized ghost imaging with two detectors}
As seen in Eq. \ref{DGIi}, larger values of $S$ measured by the bucket detector results in a greater weight for that particular speckle field, therefore external noise sources can still affect the overall reconstruction. There exists another iterative algorithm which instead normalizes each individual measurement $S$, as well as the running average, according to the reference signal $R$, resulting in an arguably more intuitive approach for dealing with time varying noise sources. We call this approach normailized GI (NGI). The algorithm used to describe each contribution to the reconstruction in NGI is given by
\begin{equation}
\label{NGI}
O_i(x,y) = \left(\frac{S}{R}-\frac{\left<S\right>}{\left<R\right>}\right)\left(I(x,y)-\left<I(x,y)\right>\right),
\end{equation}
where we have assumed $\frac{\left<S\right>}{\left<R\right>} \approx \left<\frac{S}{R}\right>$ for a large number of measurements. By considering equations \ref{DGIi} and \ref{NGI} we can summarize the difference between the two algorithms as
\begin{equation}
\label{SignalComparison}
\left<O(x,y)_{NGI}\right>=\frac{1}{\left<R\right>}\left<O(x,y)_{DGI}\right>.
\end{equation}

\subsubsection{Normalized ghost imaging with a single detector}\label{SNGI}
In a computational GI setup, we can show that the additional detector used to measure the reference signal in DGI and NGI can instead be estimated based on the known light field reflected from the SLM and the average measured signal $S$ for an arbitrary number of previous iterations. Calculating $R$ negates the requirement for an additional detector, whilst improving the performance of the reconstruction compared to TGI, thus single-detector NGI (SNGI) is identical to the TGI experimental setup, with only a modified algorithm.

\subsection{Signal-to-noise ratio analysis}

\begin{figure}[htbp]
\centering\includegraphics[width=12cm]{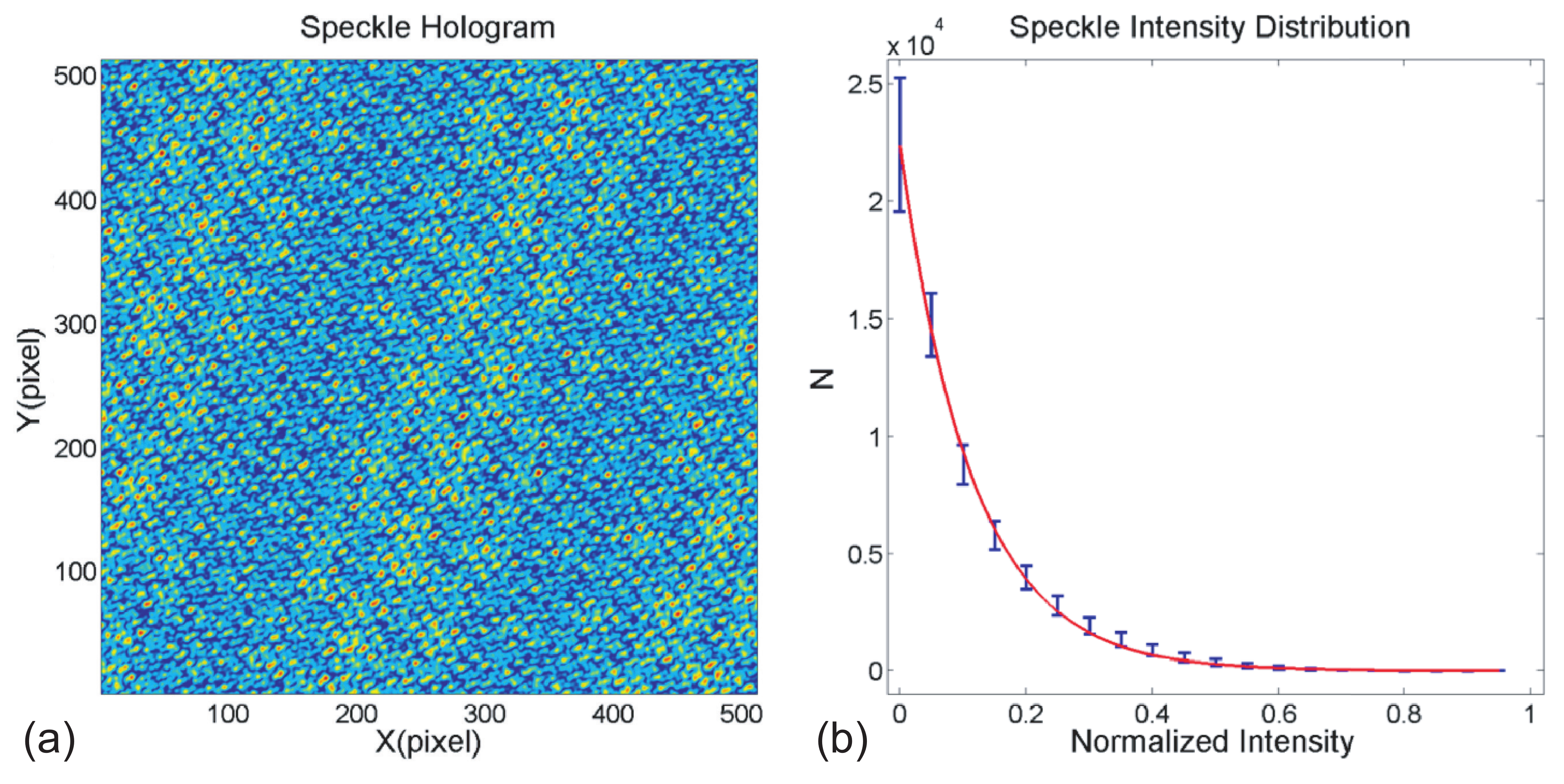}
\caption{(a) A typical speckle pattern hologram. (b) The measured intensity distribution of the speckle pattern (blue) and an exponential curve (red).}
\label{Figuretwo}
\end{figure}

To make a quantitative comparison between the NGI and the existing algorithms, we adopt a similar approach as used by Ferri {\it et al} and investigate the theoretical contribution to the signal-to-noise ratio (SNR) for objects with varying transmission functions. In \cite{Ferri2010PRL} the authors express the average quantity of Eq.\,\ref{DGIi} in terms of the object transmission fluctuation $\delta T(x,y) = T(x,y)-\overline{T}$,
\begin{equation}
\label{DGImean}
\left<O(x,y)_{DGI}\right>=\overline{A_{s}}\left<I\right>^2\delta T(x,y),
\end{equation}
where $\overline{A_{s}}$ is the average speckle area and $\overline{T}=\int_{A_{l}}\left< I(x,y)\right>T(x,y)dxdy / \int_{A_{l}}\left<I(x,y)\right>dxdy$ is the average transmission function of the object. Note that Eq. \ref{DGImean} is obtained under the assumptions of uniform illumination (the average speckle beams are constant over their area) and perfect resolution (the speckle area is much smaller compared to features of the object). The corresponding signal of DGI can be defined as
\begin{equation}
\label{signalDGI}
(\Delta \left<O_{DGI}\right>)^2=\overline{A_{s}}^2\left<I\right>^4(\Delta T)^2,
\end{equation}
where $\Delta T$ is the variation of the object transmission function to be detected. Similarly, using Eq. \ref{SignalComparison}, we can express the signal of NGI as
\begin{equation}
\label{NGIsignal}
(\Delta \left<O_{NGI}\right>)^2=\overline{A_{s}}^2\frac{\left<I\right>^4}{\left<R\right>^2}(\Delta T)^2.
\end{equation}

The speckle patterns used in our experiment exhibit complex-Gaussian behaviour, such that the intensity is exponentially distributed (see Fig.\,\ref{Figuretwo}), and the noise associated to the measurement of $O(x,y)$ can be expressed as
\begin{equation}
\label{NGIvariance}
\left<\delta O^2(x,y)\right>=\left<O(x,y)^2\right>-\left<O(x,y)\right>^2,
\end{equation}
for which it can be shown that $\left<O(x,y)\right> = 0$, thus the second term on the right hand side (RHS) of in Eq.\,\ref{NGIvariance} may be omitted. Again, under the assumptions of uniform illumination and perfect resolution, the noise of DGI can be expressed as
\begin{equation}
\left< O^2_{DGI}\right>\approx A_sA_l\left<I\right>^4\overline{\delta T^2},
\end{equation}
where $\overline{\delta T^2}=\overline{T^2}-\overline{T}^2$ and $\overline{T^2}=\int_{A_{l}}\left< I(x,y)\right>T^2(x,y) dxdy / \int_{A_{l}}\left<I(x,y)\right>dxdy$. Using linearization we can write
\begin{equation}
\frac{S}{R} \approx \frac{\left<S\right>}{\left<R\right>}\left(1+\frac{\delta S}{\left<S\right>} - \frac{\delta R}{\left<R\right>}\right),
\end{equation}
where $\delta S$ and $\delta R$ are the zero-mean deviation of $S$ and $R$, thus the noise of NGI is shown to be
\begin{equation}
\left< O^2_{NGI}\right>\approx A_sA_l\frac{\left<I\right>^4}{\left<R\right>^2}\overline{\delta T^2}.
\end{equation}

Finally, we show that the SNR contribution for NGI is
\begin{equation}
\label{SNRngi}
SNR_{NGI} = SNR_{DGI}=\frac{M}{N_{speckle}}\frac{\Delta T^2}{\overline{\delta T^2}},
\end{equation}
where $N_{s}=A_{l} / A_{s}$ is the number of speckles in the field. The SNR contribution for NGI is found to be identical to that of the DGI algorithm derived in \cite{Ferri2010PRL}. For comparison the SNR contribution for TGI was shown to be 
\begin{equation}
\label{SNRtgi}
SNR_{TGI} = \frac{M}{N_{speckle}}\frac{\Delta T^2}{\overline{T^2}}.
\end{equation}
Therefore we can examine the difference between the NGI (or DGI) and TGI algorithms by obtaining the ratio of SNR calculations, given as 
\begin{equation}
\label{SNRratio}
\frac{SNR_{NGI}}{SNR_{TGI}} = 1+ \frac{\overline{T}^2}{\overline{T^{2}} - \overline{T}^2}.
\end{equation}
As highlighted by Ferri {\it et al}, the difference is always greater than 1 and dependent only upon the variation in the object transmission function.

\section{Experiment results}

\begin{figure}[t]
\centering\includegraphics[width=12cm]{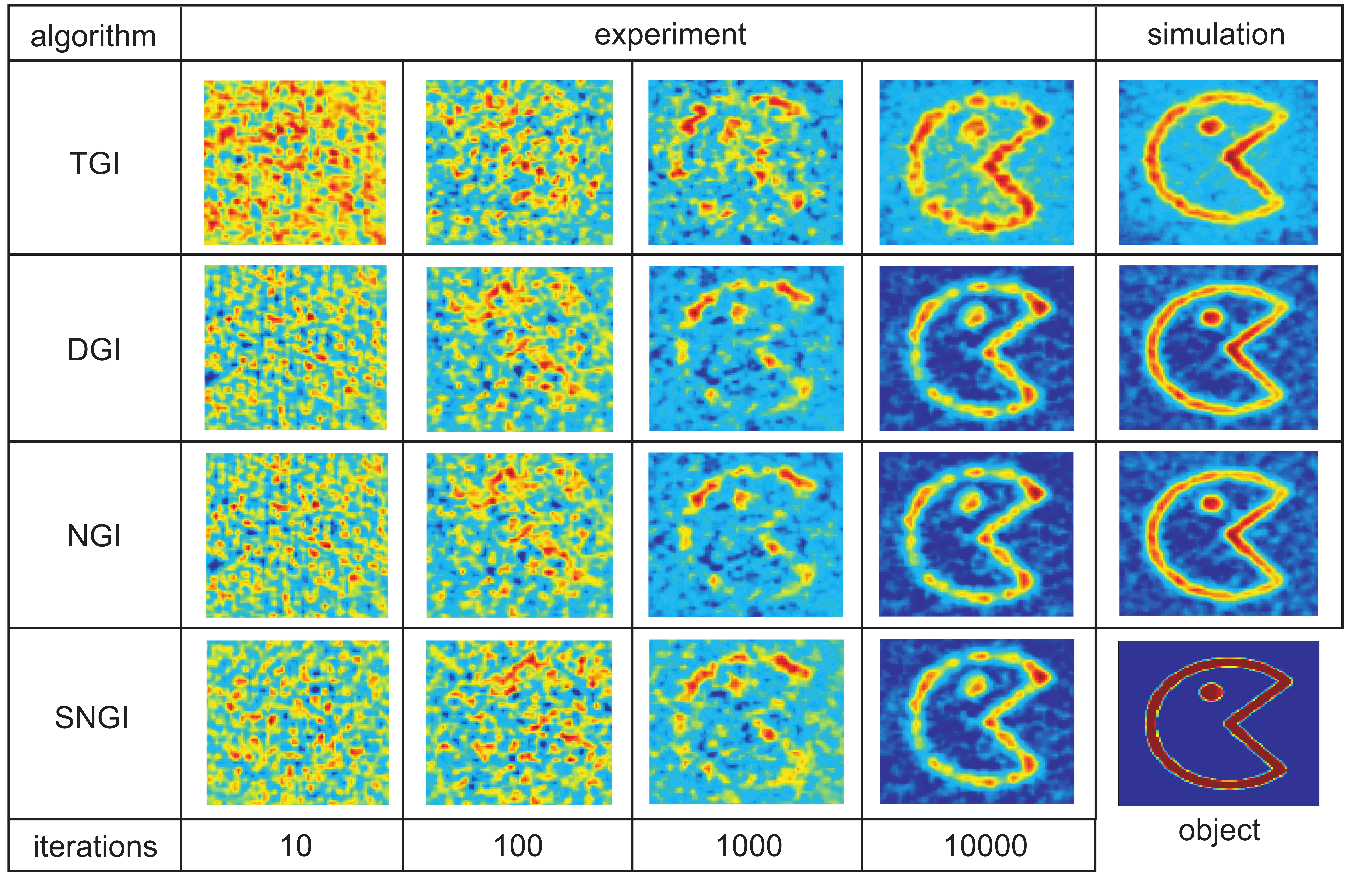}
\caption{Experimental results (middle column) for TGI, DGI and NGI reconstruction algorithms as they evolve (10, 100, 1000 and 10000 iterations from left to right, respectively) with the corresponding simulated results (right column). The transmissive object is shown in the lower right. The bottom row shows the evolution for reconstructing the object with the NGI algorithm using a single detector and predicting the reference signal $R$, termed here the SNGI algorithm.}
\label{Figurethree}
\end{figure}

We generated a series of random speckle patterns using an SLM by simulating the interference of many plane waves on a computer. The real and imaginary amplitude components and the wave vector $\vec{k}$ of each simulated plane wave is Gaussian distributed. Figure\,\ref{Figuretwo} shows a typical example of the speckle patterns generated on the SLM and the exponentially distributed intensity for many patterns, implying that the speckle hologram has complex-Gaussian statistics, thereby a good approximation for real speckle fields \cite{goodmanstatistical}. A binary transmissive object, $5{\rm mm}\times5{\rm mm}$ in size, is located after a $3\times$ magnification telescope in the image plane of the SLM. Since we know both the object and the random speckle field projected to the SLM, we are able to simulate the expected results for comparison with our experiment. Experimental and simulated reconstruction results after $10000$ iterations are shown in Fig.\,~\ref{Figurethree}. The simulated reconstruction is produced assuming no external noise sources. The partially transmissive object used is indicated in the bottom right of Fig.\,~\ref{Figurethree}. It is clear that the DGI and NGI algorithms provide very similar results, as predicted from the theory, and both show improved background subtraction compared to TGI.

Compared with the traditional computational GI setup, the NGI algorithm requires a reference bucket detector. However, as discussed in section \ref{SNGI}, the advantage of computational GI means that we can replace this bucket detector with a virtual reference detector generating a simulated $R$. Thus we can negate the requirement for the reference detector and return the system to a true single element camera, which we call single-detector NGI (SNGI). The two major factors that dominate the value of $R$ are from the different speckle patterns displayed on the SLM and fluctuations of the incident laser power. We can computationally predict changes to the value of $R$ due to the speckle pattern, whereas fluctuations of the laser power can be simulated by using a rolling average for a particular series of $S$ measurements. The bottom row in Fig.\,~\ref{Figurethree} shows the experimental results for reconstructing the object using the SNGI algorithm. We observe similar results compared with DGI and NGI algorithms indicating an improved performance compared with the TGI algorithm for single element camera.

\begin{figure}[t]
\centering\includegraphics[width=12cm]{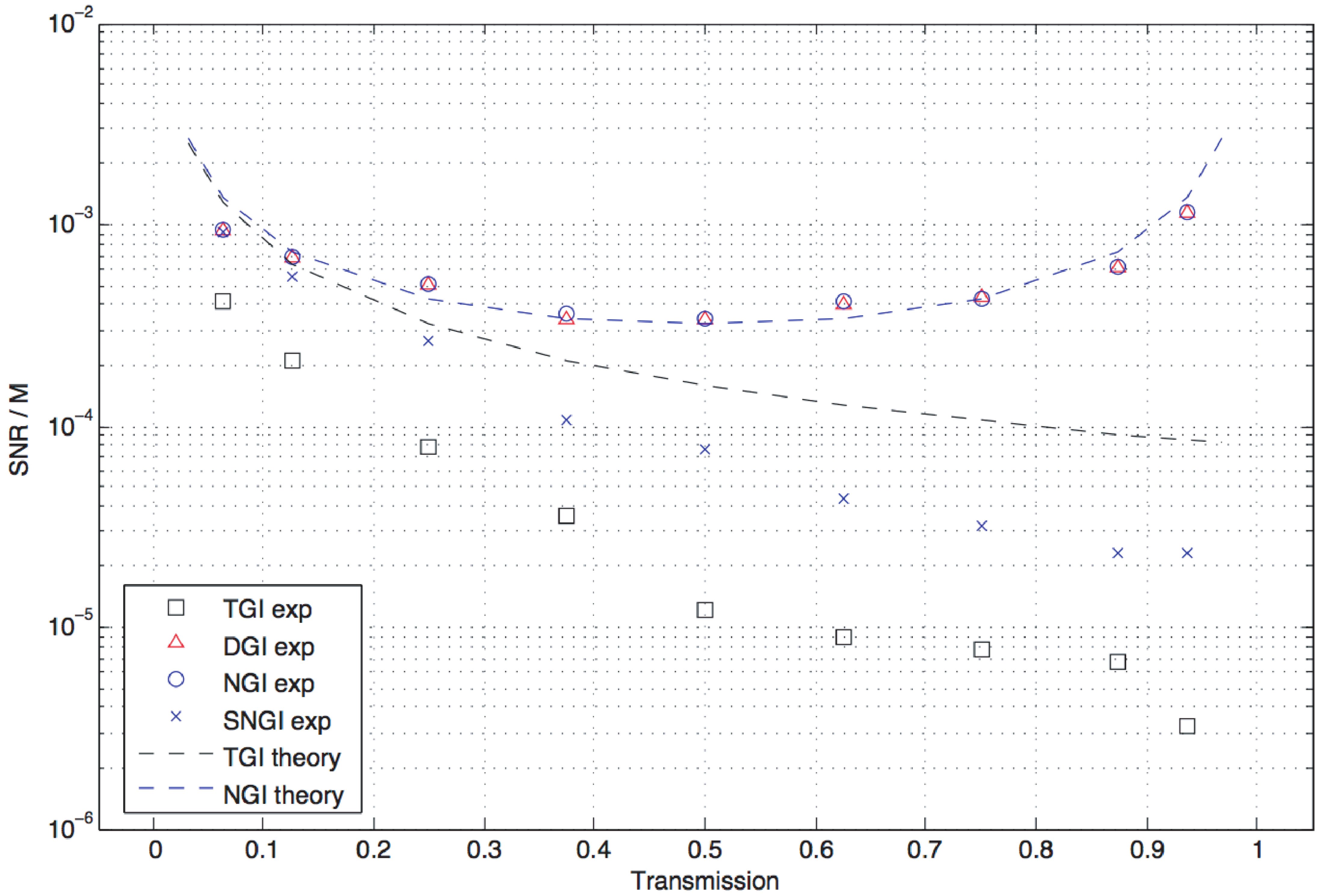}
\caption{Signal-to-noise ratio's for DGI, NGI, SNGI and TGI versus transmitting area. Transmitting ratio is defined as the ratio between the transmitting area of the object and the area of the speckle field.}
\label{Figurefour}
\end{figure}

To demonstrate the effect of object transmission function on the performance of NGI compared with TGI and DGI algorithms we used a similar experimental approach to that in Ref.\,\cite{Ferri2010PRL}. By scanning a knife edge (located in the image plane of the SLM, as before) across the speckle field in well defined steps (for which $\Delta T = 1$), we measured the SNR's for the final object reconstruction obtained after $5000$ random speckle iterations. The beam size used was $10\times10\,{\rm mm}$ and the speckle size at the plane of the object was found to be $\delta_s \sim 90\,\mu{\rm m}$, providing around $N_s \sim12500$ speckles. The experimental results and theoretical predictions for the SNR's of each iterative algorithm are shown in Fig.\,\ref{Figurefour}. Note that the $y$-axis has been normalized to the number of iterations. We observe close quantitative agreement between the theory and the measurements. The results indicate that for low transmissive objects, all algorithms reconstruct with similar SNR, while for more transmissive objects the DGI and NGI algorithms become more efficient in comparison to TGI due to the differential nature of the reconstruction. Furthermore, we observe that when using a single detector, SNGI is a more efficient algorithm for reconstructing objects of all transmissions compared to TGI. We observe that for increasing transmissive objects SNGI becomes less efficient than NGI, for which the reason is the subject of ongoing research. Similar to \cite{Ferri2010PRL}, we find a systematic discrepancy between the experimental results of TGI and the theoretical predictions.

\section{Normalization in matrix inverse algorithms}
\subsection{Introduction to matrix inverse algorithms and compressive  sensing}

As an alternative to the iterative techniques discussed above, we can choose to record all the signals for a complete set of speckle patterns and then treat the image reconstruction as one of matrix inversion. The series of $M$ speckle patterns, each containing $N$ pixels can be represented by a $M\times N$ matrix. If the object is also represented as an $N$ element column vector, then the vector containing the measured signals is a $M$ element vector. This relationship is expressed as 
 \begin{equation}
\left[\begin{array}{ccc}
S_i\\ \vdots \\S_N\end{array} \right]  = \left[ \begin{array}{ccc} \\ M \times N \\ \\ \end{array}\right] \times \left[ \begin{array}{ccc} \\ T_{(x,y)} \\ \\ \end{array}\right].
\label{Equation7}
\end{equation}

In the case where the number of speckle patterns equals the number of pixels then the $M\times N$ matrix is square, such that its inverse can be calculated and the object vector determined.  However when $M<N$ and or $N$ is large, the system is ill-conditioned and calculating the inverse of the matrix is not straightforward. Problems of this type are wide spread in physics and techniques for solving them have been developed. Within our system the appeal is to reconstruct the image of $N$ pixels from $M$ measurements where $M<N$. That this is possible is based on the fact that natural images are sparse and  the reconstruction can be obtained by solving a convex optimization problem \cite{Boyd2004}, which is a generalization of a linear least squares problem. In contrast to iterative methods, compressive GI (CGI) needs to take all measurements, represented here, in some compressible basis (in this case a discrete cosine transform which has been applied to each row of the $M\times N$ matrix). Solving the convex optimization problem requires minimizing the $\ell1$ norm \cite{Donoho2006l1}. 

\subsection{Normalized compressive ghost imaging}

By normalizing the measured object signal relative to the reference signal as performed above, such that $S' \equiv S/R$, we can apply the CGI technique \cite{Katz2009APL} to reconstruct our object. Equation\,\ref{Equation7} can then be written for normalized CGI (NCGI) as
\begin{equation}
\left[\begin{array}{ccc}
S'_i\\ \vdots \\S'_N\end{array} \right]  = \left[ \begin{array}{ccc} \\ M \times N \\ \\ \end{array}\right] \times \left[ \begin{array}{ccc} \\ T_{(x,y)} \\ \\ \end{array}\right].
\label{Equation8}
\end{equation}

\begin{figure}[t] 
   \centering
   \includegraphics[width=5in]{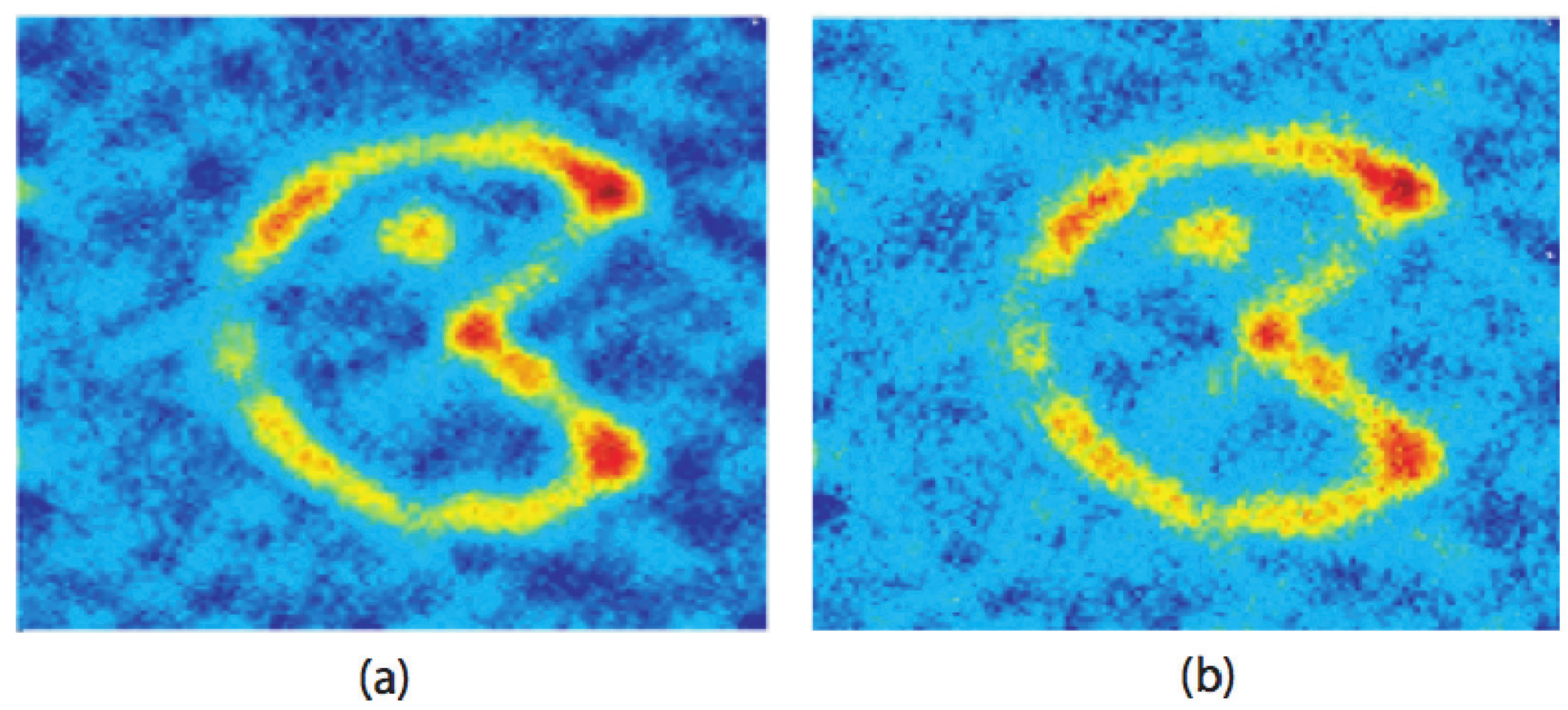} 
   \caption{(a) Experimental result of Normalized known vector reconstruction method (S/R) having SNR = 9.95. (b) Standard CGI reconstruction from S having SNR = 7.39.}
   \label{CGIfig}
\end{figure}

Performing both NCGI and CGI analyses using the same experimental data (acquired using the experimental setup in Fig.\ref{ExperimentalSetup}) we obtain the reconstruction in Fig.\,\ref{CGIfig}. We observe a clear improvement using the NCGI algorithm compared to the CGI algorithm, manifest as an increased SNR value. The efficiency with which NCGI can reconstruct sparse images over CGI is determined by the level of noise in the system. We find that when there is no system noise present, both reconstructions are essentially identical. Thus the main improvement in employing NCGI over CGI with the additional reference detector is the ability to protect the reconstruction from time varying noise sources. 

\section{Conclusion}
In conclusion we have compared different iterative GI methods to reconstruct an object and studied a new GI algorithm, which we call normalized GI (NGI). The performance of the differential GI (DGI) and NGI algorithms show good quantitative agreement as predicted by the theoretical foundations that support them. Our results indicate that by normalizing the measured signal relative to a reference signal, a more appropriate weighting factor is applied to the ensemble average of the estimated object, compared to the traditional GI (TGI) algorithm. Our analysis of the measured SNR and the object transmission shows a significant improvement for more transmissive objects in comparison to TGI. Furthermore, we have shown it is possible to apply normalization to systems with a single detector, SNGI, by estimating the reference signal. We have also investigated normalization within a compressive matrix inversion method, showing similar results to an non-normalized algorithm but with enhanced noise suppression. We believe the NGI algorithm will be a useful resource for imaging where alternative techniques are required in the future. 

\section*{Acknowledgments}
At the time when this work was ready for publication the authors found through private communication with Alessandra Gatti and Fabio Ferri that they had similar findings, for whom we thank for useful discussion. The authors thank our referee for providing constructive feedback and insight on our manuscript. MJP would like to thank the Royal Society and the Wolfson Foundation. The work of JHS was supported by the DARPA Information in a Photon (InPho) Program. We gratefully acknowledge the financial support from the UK EPSRC.
\end{document}